\documentclass{elsart}
\usepackage{natbib}
\usepackage{epsfig}

\begin{document}

\begin{frontmatter}
\title{Extragalactic Sources of TeV Gamma Rays: A Summary}
\author[sao]{D.Horan}
\author[sao]{T.C. Weekes}
\address[sao]{Whipple Observatory, Harvard-Smithsonian Center for
Astrophysics,\\ P.O. Box 97, Amado, AZ 85645-0097 USA}

\begin{abstract} 

The development of techniques whereby gamma rays of energy 100 GeV and
above can be studied from the ground, using indirect, but sensitive,
techniques has opened up a new area of high energy photon
astronomy. The most exciting result that has come from these is the
detection of highly variable fluxes of TeV gamma rays from the
relativistic jets in nearby AGN. The recent detection of signals from
a starburst galaxy and from a radio galaxy opens the possibility that
the extragalactic emission of TeV gamma rays is a ubiquitous
phenomenon. Here we attempt to summarize the properties of the sources
detected so far.

\end{abstract}


\end{frontmatter}

\section{Introduction}

This symposium comes at an interesting time in the history of Very
High Energy (VHE) gamma-ray astronomy. The TeV source catalog has now
swelled to respectable proportions and is attracting increasing
attention amongst theorists and observers at longer wavelengths. The
atmospheric Cherenkov imaging technique has been demonstrated to be an
effective tool at energies greater than 300 GeV. No other technique
has been suggested that is competitive in the energy region from 50
GeV to 50 TeV; clearly, further efforts to improve and extend the
technique are justified. In fact a new generation of instruments is
under development and the next symposium will surely be dominated by
their achievements.

It is 4.5 years since the first symposium on the TeV Astrophysics of
Extragalactic Sources was held at the CfA in Cambridge
\cite{Catanese:98}.  These proceedings show that there has been
significant progress since that time. As we discuss below, not only
has the number of detected objects increased (and with this the depth
of the universe that is probed) but the number of kinds of
extragalactic sources has also increased; this is a prediction of good
things to come as the sensitivity and range of the telescopes is
improved.

Although there has been considerable activity on the theoretical front
with the development of models to explain the observed phenomena,
there is still no consensus even on such basic things as the nature of
the progenitor particles. This is still an observation-driven field
and seems likely to be so for some time to come. Clearly the results
of multi-wavelength observations are indicative of complex mechanisms
and the simple models that have been proposed have a long way to go in
providing a full explanation.

Here we will not attempt to give either a summary of the symposium or
a comprehensive review of the field. The papers in this volume speak
for themselves and describe the progress in the field; recent reviews
can be found elsewhere \cite{Aharonian:97}; \cite{Hoffman:99};
\cite{Ong:03}; \cite{Weekes:03}. The next generation of instruments: 
CANGAROO-III, HESS, MAGIC and VERITAS will reach a level of
sensitivity such that they are effectively limited by the cosmic
electron background; they should all be online by 2004-2006.

\section{Status of TeV Gamma-ray Astronomy c. 2003}

An updated catalog of sources is shown in Table~\ref{catalog} and is
plotted in Figure 1. The criterion for inclusion in this catalog is
that the detection should be significant and be published in the
refereed literature. Four sources have been added since the last
catalog \cite{Weekes:01}. Of more significance perhaps is that three
new classes of object are represented (starburst galaxies, radio
galaxies, OB Associations).  It is noteworthy that many of the sources
listed are not in the EGRET Catalog \cite{Hartman:99}, an indication
that the TeV sky opens a new window on the universe. The allotted
grade gives some measure of the credibility that should be assigned to
the reported detections; ``A'' sources have been independently
verified at the 5 $\sigma$ level. On this scale the EGRET sources
would be classified as ``B'' (except for 3C273 which would be
classified as ``A'' since it was also detected by COS B
\cite{Swanenburg:78}; \cite{Bignami:81}). In this new catalog the
status of several of the TeV sources (marked with an asterisk) has
been upgraded (and only one has been downgraded). This suggests that
the field has achieved a degree of maturity and that systematic
effects in most experiments (but perhaps not all) are understood and
accounted for. We propose therefore that an acceptable standard for
the publication of a claim for the detection of a new source by a
mature experiment should be the 4 $\sigma$ level of significance.

\begin{table}
\caption{Source Catalog c.2003}
\label{catalog}
\begin{center}
\begin{tabular}{llllll} \hline  

TeV Catalog     & Source         &   Type     &   Discovery  & EGRET         & Grade \\
Name            &                &            &   Date/Group & 3rd. Cat.     &       \\
\hline
\\

TeV 0047$-$2518 & NGC 253        & Starburst  & 2003/CANG.   & no            & B*    \\
TeV 0219+4248   & 3C66A          & Blazar     & 1998/Crimea  & yes           & C$-$  \\
TeV 0535+2200   & Crab Nebula    & SNR        & 1989/Whipple & yes           & A     \\
TeV 0834$-$4500 & Vela           & SNR        & 1997/CANG.   & no            & C*    \\
TeV 1121$-$6037 & Cen X-3        & Binary     & 1999/Durham  & yes           & C     \\
TeV 1104+3813   & Mrk\,421       & Blazar     & 1992/Whipple & yes           & A     \\
TeV 1231+1224   & M87            & Radio Gal. & 2003/HEGRA   & no            & C     \\
TeV 1429+4240   & H1426+428      & Blazar     & 2002/Whipple & no            & A*    \\
TeV 1503$-$4157 & SN1006         & SNR        & 1997/CANG.   & no            & B*    \\
TeV 1654+3946   & Mrk\,501       & Blazar     & 1995/Whipple & no            & A     \\
TeV 1710$-$4429 & PSR 1706$-$44  & SNR        & 1995/CANG.   & no            & A     \\
TeV 1712$-$3932 & RXJ1713.7$-$39 & SNR        & 1999/CANG.   & no            & B+*   \\
TeV 2000+6509   & 1ES1959+650    & Blazar     & 1999/TA      & no            & A*    \\
TeV 2032+4131   & CygOB2         & OB assoc.  & 2002/HEGRA   & yes$^\dagger$ & B     \\
TeV 2159$-$3014 & PKS2155$-$304  & Blazar     & 1999/Durham  & yes           & A*    \\
TeV 2203+4217   & BL Lacertae    & Blazar     & 2001/Crimea  & yes           & C*    \\
TeV 2323+5849   & Cas A          & SNR        & 1999/HEGRA   & no            & B*    \\
TeV 2347+5142   & 1ES2344+514    & Blazar     & 1997/Whipple & no            & A*    \\

\hline
\end{tabular}
\end{center}

\noindent $^\dagger$ CygOB2 lies within the 95\% error ellipse of the EGRET source 
3EG J0233+4118.

\end{table}

\begin{figure}
\begin{center}
{\resizebox{\textwidth}{!}{\includegraphics[draft=false]{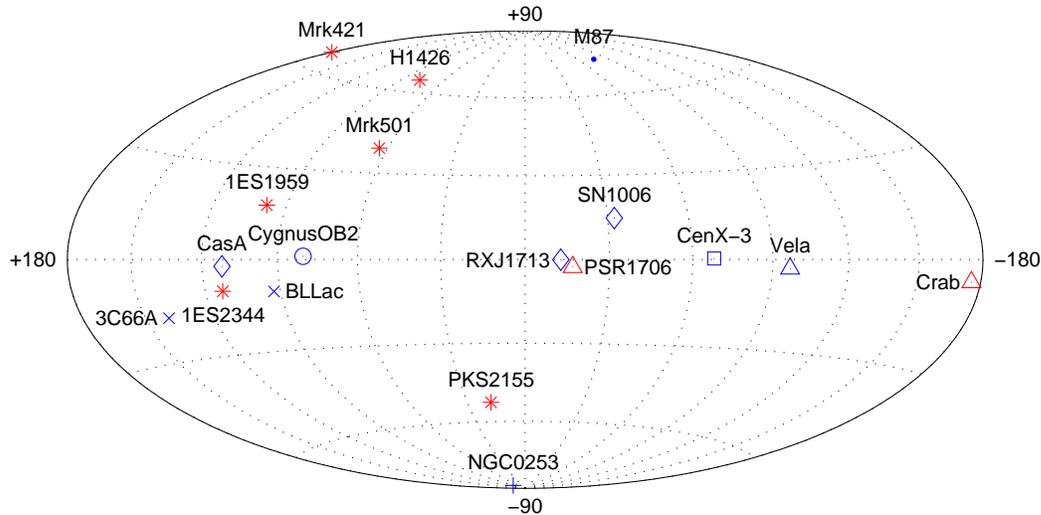}}}
\caption{The 18 claimed sources of TeV gamma rays c. 2003; confirmed
sources are drawn in red.}
\end{center}
\end{figure}

\section{Extragalactic TeV Sources}

Arguably the most exciting results to come from the search for TeV
sources of gamma rays has been the discovery that many of the nearby
blazars are detectable sources. This was certainly not anticipated
although some of the earliest TeV observations were directed towards
radio galaxies \cite{Chudakov:65} and quasars \cite{Long:65}. The
detection of 3C273 by COS B should perhaps have alerted TeV observers
to the possibility that AGN might be denizens of the TeV cosmic
zoo. However the spectrum was rather soft and did not suggest that
even this, the closest quasar, would be detectable at TeV energies. By
and large, the AGN community was equally unexcited by this detection
and the discovery by EGRET that the 100 MeV sky was dominated by
blazars came as a real surprise to the larger high energy astrophysics
community.

The fact that these EGRET AGN all had flat spectra with little sign of
a flattening above 10 GeV was auspicious for TeV astronomy.  However
no firm predictions were made of the fluxes to be expected and early
observations of a selection of these sources by ground-based observers
had disappointing results \cite{Kerrick:95}. It was not until the
detection of Markarian 421 (Mrk\,421) \cite{Punch:92} that it became
apparent that AGN might be an important TeV phenomenon.

By the time AGN had been established as MeV-GeV sources by EGRET, the
sensitive atmospheric Cherenkov imaging technique had been
developed. Given the high degree of variability now detected in TeV
sources it is perhaps fortunate that the first source seen at VHE
energies was the Crab Nebula \cite{Weekes:89}, a notoriously steady
source. One can only speculate at the controversy that would have
ensued if the highly variable Mrk\,421 had been the first source
reported with the new VHE techniques!

In all, ten extragalactic objects have been reported as sources of VHE
gamma rays and their properties are summarized in
Table~\ref{tev1}. The integral fluxes that were reported in the
detection papers (referenced in the second column), are quoted above
the peak response energy (E$_{peak}$) at which they were detected (as
given in the last column).  Their positions are shown in
Table~\ref{tev2}. It should be noted that only six of the AGN entries
have been independently confirmed at the 5 $\sigma$ level and hence
are classified as A sources.  All of the VHE blazars detected to date
are relatively close-by with redshifts ranging from 0.031 to 0.129,
and perhaps to 0.444; they are all members of the BL Lacertae blazar
subclass (BL Lacs). These objects are characterized by few or no
emission lines and are labeled Low frequency peaked (LBL) or High
frequency peaked (HBL) depending upon the waveband, radio or X-ray, in
which their detected synchrotron emission peaks.

Accurate derivation of the VHE spectrum is important for many
reasons. Since these are the most energetic photons detected from
blazars, the shape of their spectrum is an important input parameter
to emission models and can therefore impose severe constraints on
them. Multi-wavelength campaigns involving TeV observations which
provide information on how the TeV spectrum varies with flux constrain
key model input parameters. By comparing such variations to those at
longer wavelengths, especially when the observations are simultaneous,
much can be inferred about the location and the nature of the
progenitor particles. Spectral features, such as breaks or cut-offs,
can indicate changes in the primary particle distribution or
absorption of the gamma rays via pair-production with low energy
photons at the source or in intergalactic space.

The high flux VHE emission from Markarian 501 (Mrk\,501) in 1997
\cite{Aharonian:97a}; \cite{Samuelson:98}; \cite{Krennrich:99}; 
\cite{Aharonian:99} and Mrk\,421 in 2001 \cite{Krennrich:01}; 
\cite{Aharonian:02}; \cite{Krennrich:02} has permitted detailed 
spectra to be extracted. Measurements are possible over nearly two
decades of energy. As many as 25,000 photons were detected in these
outbursts so that the spectra were derived with high statistical
accuracy. Unlike the HE sources where the photon-limited blazar
measurements are consistent with a simple power law, there is definite
structure seen in the VHE measurements \cite{Krennrich:99};
\cite{Aharonian:01} with evidence for an exponential cut-off in the 
spectra of some blazars. For Mrk\,421, this cut-off is $\approx$ 4 TeV
and for Mrk\,501 it is $\approx$ 3-6 TeV. The coincidence of these two
values suggests a common origin, i.e., a cut-off in the acceleration
mechanisms in the blazars or perhaps the effect of the infra-red
absorption in extragalactic space. Attenuation of the VHE gamma rays
by pair-production with background infra-red photons could produce a
cut-off that is approximately exponential. Indeed, consistent with
this expectation, the spectrum of the more distant blazar H1426+428
shows evidence for spectral flattening at energies above 1 TeV
\cite{Aharonian:03}.

\begin{table}
\caption{Extragalactic TeV Sources: Gamma-ray Properties}
\begin{center}
\label{tev1}
\begin{tabular}{lllccc} \hline  
\multicolumn{2}{l}{Source (Det. Paper)}     & Class          & F$_\gamma$ (mean)          & F$_\gamma$ (Det.)           &  E$_{peak}$ \\
\multicolumn{2}{l}{}                        &                &  $>$ 100 MeV               & $>$ E$_{peak}$              & (Det.)      \\
\multicolumn{2}{l}{}                        &                & 10$^{-8}$cm$^{-2}$s$^{-1}$ & 10$^{-12}$cm$^{-2}$s$^{-1}$ & TeV         \\
\hline

NGC 253 & \cite{Itoh:02}                    & Starburst Gal. & U.L.                       & 7.8                         & 0.52        \\
3C66A  & \cite{Neshpor:98}                  & BL Lac(LBL)    & 18.7                       & 30.0                        & 0.90        \\
Mrk\,421 &\cite{Punch:92}                   & BL Lac(HBL)    & 13.9                       & 15.0                        & 0.50        \\
M87 & \cite{Aharonian:03a}                  & Radio Galaxy   & U.L.                       & 1.0                         & 0.73        \\
H1426+428 & \cite{Horan:02}                 & BL Lac(HBL)    & U.L.                       & 20.4                        & 0.28        \\
Mrk\,501 & \cite{Quinn:96}                  & BL Lac(HBL)    & U.L.                       & 8.1                         & 0.30        \\
1ES1959+650 & \cite{Nishiyama:00}$^\dagger$ & BL Lac(HBL)    & U.L.                       & 29.4                        & 0.60        \\
PKS2155$-$304 & \cite{Chadwick:99}            & BL Lac(HBL)    & 13.2                       & 42.0                        & 0.30        \\
BL Lacertae & \cite{Neshpor:01}             & BL Lac(LBL)    & 11.1                       & 21.0                        & 1.00        \\
1ES2344+514  &\cite{Catanese:98a}           & BL Lac(HBL)    & U.L.                       & 11.0                        & 0.35        \\
\hline
\end{tabular}
\end{center}

\noindent $^\dagger$ No flux was quoted in the initial detection paper \cite{Nishiyama:00}; 
the flux from \cite{Holder:03} is quoted here.

\end{table}

\begin{table}
\caption{Extragalactic TeV Sources: Position and Size}
\begin{center}
\label{tev2}
\begin{tabular}{lllrrrl} \hline  
Source                 & $z$            & R. A.    & Declination & Gal. Lat. & Gal. Long.     \\ 
                       &                & h/m/s    & d/m/s       & d/m/s     & d/m/s          \\
\hline

NGC 253$^\dagger$      & 0.0006 & 00 47 06 & $-$25 18 35   &  94 32 39 & $-$87 56 15      \\
3C66A                  & 0.444  & 02 19 30 &   +42 48 30   & 139 39 42 & $-$17 11 04      \\
Mrk\,421               & 0.031  & 11 04 27 &   +38 12 32   & 179 49 56 &   +65 01 50      \\
M87$^{\dagger\dagger}$ & 0.004  & 12 30 49 &   +12 23 28   & 283 46 18 &   +74 29 26      \\
H1426+428              & 0.129  & 14 28 33 &   +42 40 20   &  77 29 07 &   +64 53 53      \\
Mrk\,501               & 0.034  & 16 53 52 &   +39 45 36   &  63 35 59 &   +38 51 35      \\
1ES1959+650            & 0.048  & 19 59 59 &   +65 08 55   &  98 00 13 &   +17 40 10      \\
PKS2155$-$304          & 0.117  & 21 58 52 & $-$30 13 32   &  17 43 50 & $-$52 14 44      \\
BL Lacertae            & 0.069  & 22 02 43 &   +42 16 40   &  92 35 20 & $-$10 26 26      \\
1ES2344+514            & 0.044  & 23 47 05 &   +51 42 18   & 112 53 31 & $-$09 54 28      \\

\hline
\end{tabular}
\end{center}

\noindent $^\dagger$The emission from this object has been found to be extended.\\
\noindent $^{\dagger\dagger}$Although the results are compatible with a point-like source, extended
emission cannot be excluded.\\
\end{table}
 
\newpage

\section{Source Narrative}

{\it NGC 253: } This is the first starburst galaxy detected and also
the closest (2.5 Mpc) source of extragalactic gamma rays. Starburst
galaxies are the site of extraordinary supernovae activity and were
postulated to be sources of VHE cosmic rays and gamma rays
\cite{Volk:96}. The detection by CANGAROO-II in 2002 was at the 11
$\sigma$ level \cite{Itoh:02}. It was observed to have a very steep
spectral index ($-$3.75) which implies that most of the signal is close
to the telescope threshold. The source was extended with the same
elongation as the optical source. A model of the source has been
proposed \cite{Itoh:03}. There are many other nearby starburst
galaxies (e.g. M82, M81, IC342) so this detection opens up the
possibility that there will be many more starburst detections even
with the present generation of telescopes.

{\it 3C66A:} This is, perhaps, the least certain of the TeV
detections; it was reported by the Crimean Astrophysical Observatory
at the 5.1 $\sigma$ level of significance \cite{Neshpor:98}. Although
a well-studied AGN and detected by EGRET, 3C66A seems an unlikely
candidate for TeV emission because of its large redshift (z = 0.444)
and its classification as an LBL. Upper limits have come from other
observations \cite{Kerrick:95}; \cite{Horan:03} but these were at
other epochs and the source could be time variable on long
time-scales. It was bright at longer wavelengths during the
observations by the Crimean group.

{\it Mrk\,421:} This was the first AGN detected at TeV energies
\cite{Punch:92} and it remains the prototype of TeV AGN because its
signal strength is, on average, 30\% of the Crab (and often much
stronger). It is the weakest blazar detected by EGRET in the 3rd
Catalog \cite{Hartman:99} and also the closest.

Whipple observations of Mrk\,421 during 1994 revealed the first clear
detection of flaring activity in the VHE emission from an AGN. A
10-fold increase in the flux, from an average level that year of
approximately 15\% of the Crab flux to approximately 150\% of the Crab
flux was observed. The observations of Mrk\,421 in 1995
\cite{Buckley:96} revealed several distinct episodes of flaring
activity as in previous observations; perhaps more importantly though,
they indicated that the VHE emission from Mrk\,421 was best
characterized by a succession of day-scale or shorter flares with a
baseline emission level below the sensitivity limit of the Whipple
detector.

The VHE emission from Mrk\,421 was seen to flare on sub-day
time-scales in 1996, with the observations of two short flares
\cite{Gaidos:96}. In the first flare the flux increased monotonically
during the course of $\sim$2 hours of observations. This flux is the
highest observed from any VHE source to date. The second flare,
observed a week later, although weaker, was remarkable for its very
short duration: The entire flare lasted approximately 30 minutes with
a doubling and decay time of less than 15 minutes.  These two flares
exhibited the fastest time-scale variability, by far, seen from any
blazar at any gamma-ray energy.

During 2001, Mrk\,421 exhibited exceptionally strong and long-lasting
flaring activity \cite{Jordan:01}. It was observed extensively with
the Whipple telescope during this time and a large database of over
23,000 gamma-ray photons was collected allowing very accurate
spectral information to be derived \cite{Krennrich:01}. The data are
best described by a power law with an exponential cutoff:

${{\rm dN}\over{\rm dE}} \propto {\rm E}^{-2.14\pm 0.03_{\rm stat}}
\exp\left[-{{\rm E}\over{4.3 \pm 0.3_{\rm stat}}}\right]$ m$^{-2}$ s$^{-1}$ TeV$^{-1}$ where E is in units of TeV.

The data were binned according to the flux level that Mrk\,421 was in
when they were recorded and spectra were derived for each flux
level. A clear correlation was found to exist between the spectral
index and the photon flux and, the spectra were all found to have
exponential cutoffs consistent with the average value of 4.3 TeV
\cite{Krennrich:02}. Spectral measurements of Mrk\,421 from previous
observing seasons by the Whipple collaboration were also found to be
consistent with this flux-spectral index correlation, suggesting this
to be a long-term property of the source. The spectral index was found
to vary between $1.89 \pm 0.04_{stat} \pm 0.05_{syst}$ in a high flux
state and $2.72 \pm 0.11_{stat} \pm 0.05_{syst}$ in a low state.

During its dramatic outburst in 2001, Mrk\,421 was also detected by
the STACEE detector at energies above 140 GeV \cite{Boone:02} and by
the CANGAROO-II telescope at energies above 10 TeV \cite{Okumura:02}.

{\it M87:} This is one of the brightest nearby radio galaxies and is
an obvious potential source of high energy radiation since the jet
displays evidence for synchrotron radiation and time variability.  The
angle of the jet is about 30$^\circ$ \cite{Bicknell:96} which means
that it is unlikely to have the same observational gamma-ray
properties as the blazars. In fact it was not detected by EGRET and
the positive observation by the HEGRA group \cite{Aharonian:03a} was a
surprise.  Although the detection was only at the 4 $\sigma$ level of
significance (weaker than any of the other sources in the TeV catalog)
it is a potentially exciting result as it opens up the possibility
that many AGN may be observable whose axes are not pointing directly
towards us. It is a weak source and its detection required 83 hours of
observation. It was not seen in observations at lower energies
\cite{LeBohec:01}; \cite{LeBohec:03} but the exposures, and hence the
flux sensitivities, were limited. The detection of M87 revives
interest in the reported detection of Centaurus A in 1975
\cite{Grindlay:75} which, although not confirmed in later, more
sensitive, observations, was at a time when the source had an
abnormally high microwave flux.

{\it H1426+428:} This source is of interest primarily because, at a
redshift of 0.129, it is the most distant confirmed source of TeV
gamma rays; three different groups have reported significant
detections \cite{Horan:01}; \cite{Aharonian:02a};
\cite{Djannati:03}. It is weak source (typically 6\% of the Crab) 
and, having its synchrotron peak located at higher frequencies than
any of the other TeV blazars, is classified as an ``extreme'' HBL
\cite{Costamante:01}. It was not seen by EGRET. The initial detection
by the VERITAS group was at the 5.8 $\sigma$ level of significance and
was based on 44.4 hours of observation. The source is also significant
in that it was predicted to be a detectable TeV emitter based on its
hard X-ray spectrum \cite{Costamante:02}. The source is definitely
variable on time-scales of a year and maybe on times as short as a
day. The energy spectrum, which has been derived by three groups, is
found to be quite steep. It is well described by a power law with a
spectral index between 250 GeV and 1 TeV of $3.50 \pm 0.15$ derived by
the VERITAS Collaboration \cite{Petry:02} and of $3.66 \pm 0.41$ by
CAT \cite{Djannati:03}. The HEGRA group derived a spectral index of
$2.6 \pm 0.6$ between 700 GeV 1.4 TeV; above this energy, consistent
with the expected signature of absorption of the TeV gamma rays by the
extragalactic infra-red photons, evidence for a break in the spectrum
was found \cite{Aharonian:03}.

{\it Mrk\,501:} Historically Mrk\,501 is important because, when it
was detected in 1995 \cite{Quinn:96}, it was the first TeV source to
be detected that had not previously been detected by EGRET; hence it
established TeV extragalactic astronomy as a discipline in its own
right. The properties of Mrk\,501 are very similar to those of
Mrk\,421 although in general the characteristic time-scales seem
longer with the flux levels varying less rapidly.  In 1997, the VHE
emission from Mrk\,501 increased dramatically.  Fortunately this was a
time when several new telescopes were coming on-line so that it was
well-observed \cite{Bradbury:97}; \cite{Djannati:99}.  After being the
weakest known source in the VHE sky, in 1995-96 it became the
brightest, with an average flux greater than that of the Crab Nebula
(whereas previous observations had never revealed a flux $>$50\% of
the Crab flux).  The amount of day-scale flaring increased and, for
the first time, significant hour-scale variations were seen. It was
also detected by EGRET for the first time. The Mrk\,501 spectrum is
similar to that of Mrk\,421 and can be represented by:

${{\rm dN}\over{\rm dE}} \propto {\rm E}^{-1.92\pm 0.03_{\rm stat}
\pm 0.20_{\rm syst}} \exp\left[-{{\rm E}\over{6.2 \pm 0.4_{\rm stat}
{_{-1.5}^{+2.9}}_{\rm syst}}}\right]$ m$^{-2}$ s$^{-1}$ TeV$^{-1}$

\cite{Aharonian:99} where E is in units of TeV.

{\it 1ES1959+650:} This was first reported in conference proceedings
by the Telescope Array group operating in Dugway, Utah
\cite{Nishiyama:00} in 2000.  Upper limits were reported by other
groups \cite{Catanese:97}. In 2002, the HEGRA Collaboration reported
their detection of 1ES1959+650 \cite{Konopelko:02}. The detection was
dramatically confirmed later in 2002 when an outburst was seen by
several groups \cite{Holder:03}; \cite{Aharonian:03b};
\cite{Djannati:03a}. Although the quiescent level was only about 5\%
of the Crab, when it flared its flux was 5 times that of the
Crab. Correlations were reported with optical and X-ray observations
\cite{Schroedter:02}. 1ES1959+650 was not seen by EGRET. The
differential energy spectrum has been derived by Aharonian et
al.\cite{Aharonian:03b} and is well described by a power law with an
exponential cut-off during flaring states:

${{\rm dN}\over{\rm dE}} \propto{\rm E}^{-1.83\pm 0.15_{\rm stat} \pm 0.08_{syst}}
\exp\left[-{{\rm E}\over{4.2{^{+1.8}_{-0.6}}_{stat} \pm 0.9_{syst}}}\right]$ m$^{-2}$ s$^{-1}$ TeV$^{-1}$ where E is in units of TeV.

The low state spectrum is best represented by a pure power law
\cite{Aharonian:03b} with a spectral index of $3.18 \pm 0.17_{stat} 
\pm 0.08_{syst}$.

Extensive multiwavelength observations on 1ES1959+650 were carried out
during May-July 2002 \cite{Holder:03a}. During this time, a TeV
gamma-ray flare was observed which had no counterpart at X-ray
energies. This ``orphan flare'' is difficult to explain in terms of
one-zone synchrotron self-Compton (SSC) models. Several possibilities
are explored in \cite{Krawczynski:03} including multiple-component SSC
models, external Compton models and proton models. The latter seems
the least likely explanation since the X-ray and gamma-ray flux from
1ES1959+650 were found to be correlated during the rest of the
observing campaign.

{\it PKS2155$-$304:} This BL Lac is very bright in the ultraviolet and
is classified as a HBL. It is highly variable in X-rays. It is
strongly detected by EGRET with a hard spectrum (index
$-1.71\pm0.24$). It was first reported by the University of Durham
group working in Narrabri, Australia \cite{Chadwick:99} who saw a
signal at the 6.8$\sigma$ level of significance. Upper limits were
also reported \cite{Chadwick:99a}; \cite{Roberts:99}. It was
dramatically confirmed by the report at this symposium by the HESS
group who saw it at the 10.1$\sigma$ level in just 2.2 hours of
observation. This is the first blazar detected in TeV gamma rays in
the southern hemisphere.

{\it BL Lacertae:} 

This is the object after which this class of AGN is named. It is now
classified as a LBL like many of the EGRET-detected AGN. The paper
\cite{Catanese:97a} that reported the detection of BL Lacertae by EGRET also
reported an upper limit at TeV energies from the Whipple
group. Subsequently the Crimean group reported the detection of this
source at the 7.2 $\sigma$ level of significance \cite{Neshpor:01}.
It was optically quite bright at this epoch (July-September, 1998). BL
Lac and 3C66A are the only LBLs that have been reported at TeV
energies; it is important that these detections at TeV energies be
confirmed as they place severe constraints on source models.

{\it 1ES2344+514:} This AGN was reported by the Whipple group in 1998
\cite{Catanese:98a} as a TeV source based on observations made in
1995. Most of the reported signal came in one night so, if real, the
source is highly variable. Indeed, {\it Beppo}SAX observations of
1ES2344+514 have revealed it to be highly variable at hard X-ray
energies \cite{Giommi:00} with the overall X-ray spectral shape
varying with intensity. Its synchrotron peak frequency was seen to
shift by a factor of $\approx$ 30 between observations taken in 1996
and in 1998. This behaviour is typical of HBLs and has, for example,
been also observed in Mrk\,501 \cite{Pian:98}. A confirmation of
1ES2344+514 has been reported by the HEGRA group \cite{Tluczykont:03}
and by the Whipple group \cite{Badran:01}. No spectrum has been
reported. At the peak of its flaring activity it was 60\% of the
strength of the Crab.

\section{Future Prospects}

With the next generation of ground-based arrays of telescopes now
coming on-line, it is expected that the number of extragalactic
sources detected above 100 GeV will increase tenfold.  Spectral
measurements will permit detailed models to be confronted with
observations. It may be possible to measure the spectral cut-offs
between 10 GeV and 100 GeV and to distinguish between those that are
intrinsic to the source and those that are due to the extragalactic
infra-red background.

The most exciting aspect of the recent results is the diversity of
objects that are now proving to be VHE gamma-ray sources. Hopefully
each of these detections is only the tip of the iceberg of each class
of source (Figure 2) and, as the flux sensitivity improves, the other
members of the class will be detectable. It is noteworthy that
although the number of sources in the TeV Catalog (Table 1) is still
small, the diversity of objects is large and already exceeds that of
the MeV-GeV catalogs.

\begin{figure}
\begin{center}
{\resizebox{\textwidth}{!}{\includegraphics[draft=false]{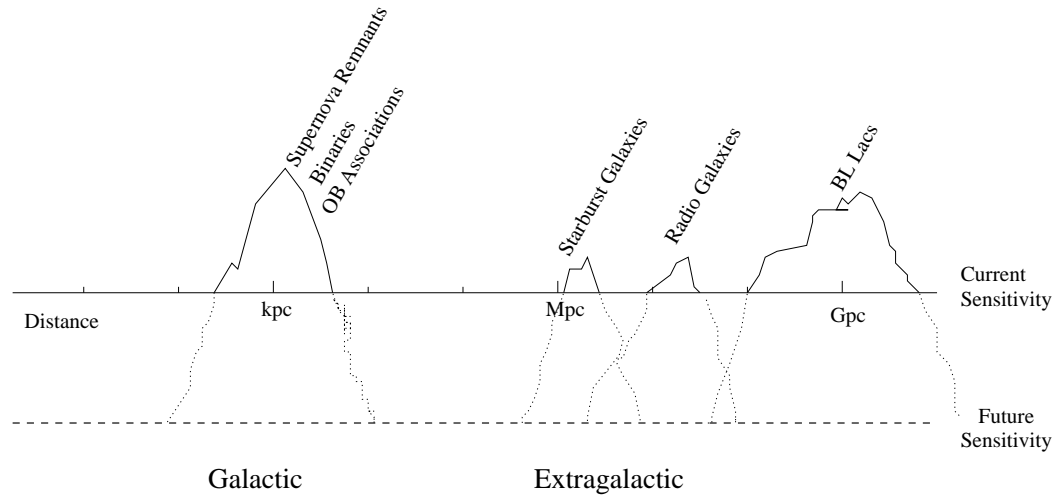}}}
\caption{Tips of the Icebergs in the TeV Universe.}
\end{center}
\end{figure}

\end{document}